\documentclass[conference,a4paper]{IEEEtran}
\usepackage{color}
\usepackage{cite}
\usepackage{amsmath}
\usepackage{amsfonts}
\usepackage{amssymb}
\usepackage{algorithm}
\usepackage{algorithmic}
\usepackage{graphicx}
\usepackage{verbatim}

\newtheorem{theorem}{Theorem}
\newtheorem{lemma}{Lemma}
\newtheorem{remark}{Remark}

\newtheorem{corollary}{Corollary}
\newtheorem{example}{Example}

\newcommand{\beq}{\begin{equation}}
\newcommand{\eeq}{\end{equation}}
\newcommand{\beqnn}{\begin{equation*}}
\newcommand{\eeqnn}{\end{equation*}}
\newcommand{\beqy}{\begin{eqnarray}}
\newcommand{\eeqy}{\end{eqnarray}}
\newcommand{\beqynn}{\begin{eqnarray*}}
\newcommand{\eeqynn}{\end{eqnarray*}}
\newcommand{\bit}{\begin{itemize}}
\newcommand{\eit}{\end{itemize}}
\newcommand{\ben}{\begin{enumerate}}
\newcommand{\een}{\end{enumerate}}
\newcommand{\bex}{\begin{example}}
\newcommand{\eex}{\end{example}}

\newcommand{\balg}[1]{\begin{algorithm} \caption{#1}}
\newcommand{\ealg}{\end{algorithm}}

\newcommand{\balgc}{\begin{algorithmic}[1]}
\newcommand{\ealgc}{\end{algorithmic}}

\newcommand{\bary}{\begin{array}}
\newcommand{\eary}{\end{array}}
\newcommand{\bmx}{\begin{bmatrix}}
\newcommand{\emx}{\end{bmatrix}}
\newcommand{\bsmx}{\left[\begin{smallmatrix}}
\newcommand{\esmx}{\end{smallmatrix}\right]}
\newcommand{\bmxc}[1]{\left[\begin{array}{@{}#1@{}}}
\newcommand{\emxc}{\end{array}\right]}
\newcommand{\bcn}{\begin{center}}
\newcommand{\ecn}{\end{center}}

\newcommand{\Rbb}{{\mathbb{R}}}
\newcommand{\Zbb}{{\mathbb{Z}}}

\newcommand{\Rnbn}{\Rbb^{n \times n}}

\newcommand{\Rmbm}{\Rbb^{m \times m}}

\newcommand{\Zn}{\Zbb^{n}}

\newcommand{\sOB}{{\scriptscriptstyle \text{OB}}}
\newcommand{\sOR}{{\scriptscriptstyle \text{OR}}}

\newcommand{\sD}{{\scriptscriptstyle \text{D}}}
\newcommand{\sR}{{\scriptscriptstyle \text{R}}}
\newcommand{\sBB}{{\scriptscriptstyle \text{BB}}}
\newcommand{\sBR}{{\scriptscriptstyle \text{BR}}}

\newcommand{\A}{\boldsymbol{A}}

\newcommand{\I}{\boldsymbol{I}}

\newcommand{\Q}{\boldsymbol{Q}}
\newcommand{\R}{\boldsymbol{R}}

\newcommand{\T}{\boldsymbol{T}}

\renewcommand{\d}{\boldsymbol{d}}

\renewcommand{\l}{\boldsymbol{\ell}}

\renewcommand{\u}{\boldsymbol{u}}
\renewcommand{\v}{\boldsymbol{v}}

\newcommand{\x}{{\boldsymbol{x}}}
\newcommand{\y}{{\boldsymbol{y}}}

\newcommand{\0}{{\boldsymbol{0}}}

\newcommand{\bx}{{\bar{x}}}

\newcommand{\bbx}{{\bar{\x}}}

\newcommand{\tv}{{\tilde{v}}}
\newcommand{\ty}{{\tilde{y}}}

\newcommand{\tby}{{\tilde{\y}}}

\newcommand{\hx}{{\hat{x}}}

\newcommand{\hbx}{{\hat{\x}}}

\newcommand{\bxi}{\boldsymbol{\xi}}

\usepackage{slashbox}

\begin{document}

\title{On the Success Probability of the Box-Constrained Rounding and Babai Detectors}


\author{\IEEEauthorblockN{Jinming~Wen\IEEEauthorrefmark{1},
Xiao-Wen~Chang\IEEEauthorrefmark{2}  and Chintha Tellambura\IEEEauthorrefmark{1}}
\IEEEauthorblockA{\IEEEauthorrefmark{1}
Department of Electrical and Computer Engineering, University of Alberta, Edmonton T6G 2V4, \\
Canada (E-mail: jinming1@ualberta.ca, chintha@ece.ualberta.ca)}
\IEEEauthorblockA{\IEEEauthorrefmark{2}School of Computer Science, McGill University,
Montreal, H3A 0E9, Canada (E-mail: chang@cs.mcgill.ca)}
}

\maketitle

\begin{abstract}
In communications, one frequently needs to detect
a parameter vector $\hbx$ in a box
from a linear model.
The box-constrained rounding detector $\x^\sBR$ and Babai detector $\x^\sBB$ are
often used to detect $\hbx$
due to their high probability of correct detection, which is referred to as success probability,
and their high efficiency of implimentation.
It is generally believed that the success probability $P^\sBR$ of $\x^\sBR$ is not larger than
the success probability $P^\sBB$ of $\x^\sBB$.
In this paper, we first present formulas for $P^\sBR$ and $P^\sBB$ for two different situations: $\hbx$ is  deterministic
and $\hbx$ is uniformly distributed over the constraint box.
Then, we give a simple example to show  that
$P^\sBR$ may be strictly larger than $P^\sBB$ if $\hbx$ is deterministic,
while we rigorously show that $P^\sBR\leq P^\sBB$ always holds
if $\hbx$ is uniformly distributed over the constraint box.
\end{abstract}


\section{Introduction}
\label{s:introduction}
Suppose that we have the following box-constrained linear model:
\begin{align}
& \y=\A\hbx+\v, \quad \v \sim \mathcal{N}(\boldsymbol{0},\sigma^2 \I) \label{e:model}, \\
& \hbx\in \mathcal{B} \equiv \{\x \in \mathbb{Z}^{n}:\l \leq \x\leq \u,\; \l, \u\in \mathbb{Z}^{n}\},
\label{e:box}
\end{align}
where $\y\in \mathbb{R}^m$ is an observation vector,
$\A\in\mathbb{R}^{m\times n}$ is a deterministic full column rank model matrix,
$\hbx$ which can be deterministic or random is an unknown integer parameter vector in the box $\mathcal{B}$,
$\v\in \mathbb{R}^m$ is a noise vector following the Gaussian distribution
$\mathcal{N}(\boldsymbol{0},\sigma^2 \I)$ with given $\sigma$.
This model arises from lots of  applications including wireless communications, see e.g., \cite{DamGC03}.

Since $\v\sim \mathcal{N}(\boldsymbol{0},\sigma^2 \I)$,
a commonly used method to detect $\hbx$ is to solve the following box-constrained integer least squares (BILS) problem:
\beq
\label{e:BILS}
\min_{\x\in\mathcal{B}}\|\y-\A\x\|_2^2,
\eeq
whose solution is the maximum likelihood detector of $\hbx$.

If $\hbx \in \Zn$ in \eqref{e:model} is not subject to any constraint, then
\eqref{e:model} is called as an ordinary linear model.
In this case, to get the maximum likelihood estimator of $\hbx$,
we solve the following ordinary integer least squares (OILS) problem:
\beq
\label{e:OILS}
\min_{\x\in\mathbb{Z}^n}\|\y-\A\x\|_2^2.
\eeq

In communications, one of the widely used methods for solving \eqref{e:BILS} and \eqref{e:OILS}
 is sphere decoding.
Although some column reordering strategies, such as  V-BLAST \cite{FosGVW99},
SQRD \cite{WubBRKK01},
and those proposed in \cite{ChaH08}  which use not only  the information of $\A$,
but also the information of $\y$ and $\mathcal{B}$, can usually reduce the computational cost
of solving \eqref{e:BILS} by sphere decoding,
solving \eqref{e:BILS} is still time-consuming especially when $\A$ is ill conditioned, $\sigma$ or $n$ is large \cite{JalO05}.
Moreover, it has been shown in \cite{Mic01} that \eqref{e:OILS} is an NP-hard problem.
Therefore, in practical applications, especially for some real-time applications,
a suboptimal detector, which can be obtained efficiently, is  often used to detect $\hbx$
instead of solving \eqref{e:BILS} or \eqref{e:OILS} to get the optimal detector.

For the OILS problem, the ordinary rounding detector $\x^\sOR$
and the Babai detector $\x^\sOB$,
which are respectively obtained by the Babai rounding off and nearest plane algorithms \cite{Bab86},
are often used suboptimal detectors for $\hbx$.
By taking the box constraint \eqref{e:box} into account, one can easily modify the algorithms for $\x^\sOR$ and $\x^\sOB$ to get box-constrained rounding detectors $\x^\sBR$
and box-constrained Babai detectors $\x^\sBB$ for $\hbx$ satisfying
both \eqref{e:model} and \eqref{e:box}.
Rounding and Babai detectors are respectively
the outputs of zero-forcing and successive interference cancellation decoders
which are widely used suboptimal detection algorithms in communications.

To characterize how good  a detector is, we use its success probability, i.e.,
the probability of the detector being equal to $\hbx$, see e.g., \cite{ChaWX13,WenC17}.

For the estimation of $\hbx$ in the ordinary linear model \eqref{e:model},
the formulas of the success probability $P^\sOR$ of the rounding detector $\x^\sOR$
and the success probability $P^\sOB$ of the Babai detector  $\x^\sOB$
have been given in \cite{WenTB16} and \cite{ChaWX13}, respectively.
Equivalent formulas of $P^\sOR$ and $P^\sOB$ were given earlier in \cite{Teu98a},
which considers the OILS problem in different formats in the application of GPS.
It is shown in \cite{Teu98a} that  $P^\sOR\leq P^\sOB$.

For the detection of $\hbx$ satisfying both \eqref{e:model} and \eqref{e:box},
it is also generally believed that
the success probability $P^\sBR$ of the rounding detector $\x^\sBR$ is not larger than
the success probability $P^\sBB$ of the Babai detector  $\x^\sBB$.

In this paper,  we develop formulas for the success probability $P_\sD^\sBR$ and $P_\sR^\sBR$ of
$\x^\sBR$ which respectively corresponding to the case that $\hbx$
is a deterministic parameter vector and  $\hbx$ is uniformly distributed over $\mathcal{B}$.
We also give a formula for the success probability $P_\sD^\sBB$ of the box-constrained Babai   detector
$\x^\sBB$ for the case that $\hbx$ is  deterministic.
Note that the success probability $P_\sR^\sBB$ of $\x^\sBB$
for the case that $\hbx$ is uniformly distributed over $\mathcal{B}$ has been given in \cite{WenC17}.
We would like to point out that the assumption that $\hbx$ follows the uniformly distribution
is often made for MIMO applications, see, e.g., \cite{JalO05}.
For the deterministic  case,  we give a simple example to show that $P_\sD^\sBR>P_\sD^\sBB$,
contrary to what we have suspected.
For the uniform random case,  however, we rigorously show that the common
belief is indeed true, i.e., $P_\sR^\sBR\leq P_\sR^\sBB$.

The rest of the paper is organized as follows.
In Section \ref{s:PBRBB}, we present formulas for $P_\sD^\sBR$, $P_\sR^\sBR$, $P_\sD^\sBB$ and $P_\sR^\sBB$.
In Section \ref{s:PRPBrelation}, we study the relationship between $P_\sD^\sBR$ and $P_\sD^\sBB$,
and rigorously show that $P_\sR^\sBR\leq P_\sR^\sBB$.
In Section \ref{s:sim}, we do simulation tests to illustrate our main results.
Finally  we summarize this paper in Section \ref{s:sum}.

{\bf Notation}. Throughout this paper, 
for $\x\in \mathbb{R}^n$, we use $\lfloor \x\rceil$ to denote its nearest integer vector, i.e.,
each entry of $\x$ is rounded to its nearest integer (if there is a tie, the one with smaller magnitude is chosen).
For a vector $\x$, $\x_{i:j}$ denotes the subvector of $\x$ formed by entries $i, i+1, \ldots,j$.
For a matrix $\A$, $\A_{i:j,i:j}$ denotes the submatrix of $\A$ formed by rows and columns $i, i+1, \ldots,j$.

\section{Success probability of box-constrained rounding and  Babai detectors}
\label{s:PBRBB}


In this section, we derive formulas for $P_\sD^\sBR$, $P_\sR^\sBR$ and $P_\sD^\sBB$.
Note that the formula for $P_\sR^\sBB$ has been derived in \cite[Th.1]{WenC17}.

Let $\A$ in \eqref{e:model} have the QR factorization
\beq
\A=[\Q_1, \Q_2]\bmx\R \\ \0 \emx,
\label{e:qr}
\eeq
where $[\underset{n}{\Q_1}, \underset{m-n}{\Q_2}]\in \Rmbm$ is orthogonal
and $\R\in \Rnbn$ is upper triangular.
Without loss of generality, we assume that  $r_{ii}>0$ for $i=1, \ldots,n$ throughout the paper.

Define $\tby=\Q_1^T\y$ and $\tilde{\v} = \Q_1^T \v$.
Then, left multiplying both sides of \eqref{e:model} with $\Q_1^T$ yields
\beq
\label{e:modelqr}
\tby=\R\hbx+\tilde{\v}, \quad \tilde{\v} \sim \mathcal{N}(\0, \sigma^2 \I).
\eeq

Let $\d=\R^{-1}\tby$,
then the box-constrained rounding detector $\x^\sBR$ and box-constrained Babai detector $\x^\sBB$
of \eqref{e:modelqr} can be respectively computed as follows:
\beq
\label{e:Rounding}
\begin{split}
x_i^\sBR=
\begin{cases}
\ell_i, \quad \mbox{ if } \lfloor d_i\rceil\leq \ell_i,\\
\lfloor d_i\rceil, \mbox{ if }  \ell_i<\lfloor d_i\rceil< u_i,\quad i=1, \ldots, n \\
u_i,\quad \mbox{ if } \lfloor d_i\rceil \geq u_i,
\end{cases}
\end{split}
\eeq
and
\beq \label{e:Babai}
\begin{split}
&  c_{i}=(\ty_{i}-\sum_{j=i+1}^nr_{ij}x_j^\sBB)/r_{ii},\\
& x_i^\sBB=
\begin{cases}
\ell_i, & \mbox{ if }\   \lfloor c_i\rceil\leq \ell_i,\\
\lfloor c_i\rceil, & \mbox{ if }\    \ell_i<\lfloor c_i\rceil< u_i,\,\; i=n,  \ldots, 1.\\
u_i, & \mbox{ if }\    \lfloor c_i\rceil \geq u_i,
\end{cases}
\end{split}
\eeq
where $\sum_{j=n+1}^nr_{nj}x_j^\sBB=0$.

\subsection{Success probability of the box-constrained rounding detector}
\label{ss:PBR}

In this subsection, we develop formulas for $P_\sD^\sBR$ and $P_\sR^\sBR$.
Since $P_\sD^\sBR$ depends on the position of $\hbx$ in the box $\mathcal{B}$,
we also give a lower bound on $P_\sD^\sBR$.
\begin{theorem}
\label{t:PBRD}
Let $\hbx$ in \eqref{e:model} be a deterministic vector, then
\beq
\label{e:PBRD}
P_\sD^\sBR=\frac{\det(\R)}{(2\pi \sigma^2)^{n/2}}\int_{I_n}\cdots \int_{I_1}
\exp\left(-\frac{\|\R\bxi\|^2_2}{2\sigma^2}\right)d\xi_1\cdots d\xi_n,
\eeq
where
\begin{equation}
\label{e:intervel I}
I_i=
\begin{cases}
(-\infty, \frac{1}{2}], \quad\mbox{ if } \hx_i=\ell_i \quad \\
[-\frac{1}{2}, \frac{1}{2}], \quad \, \;\mbox{ if } \ell_i<\hx_i< u_i.\\
[-\frac{1}{2}, \infty), \quad \mbox{ if } \hx_i=u_i
\end{cases}
\end{equation}
\end{theorem}
{\bf Proof}. Since $\hbx$ is deterministic and $\tilde{\v}\sim \mathcal{N}(\0,\sigma^2 \I)$,
by \eqref{e:modelqr}, we have
\[
\d -\hbx =\R^{-1}\tby -\hbx=\R^{-1}\tilde{\v} \sim \mathcal{N}(\0,\sigma^2(\R^T\R)^{-1}).
\]

By the definition of $\lfloor\x\rceil$, \eqref{e:Rounding} and \eqref{e:intervel I},
\[
\x^\sBR=\hbx\Leftrightarrow d_i-\hat{x}_i\in I_i, \ \ i=1,\ldots, n.
\]
Therefore, \eqref{e:PBRD} holds.
\ \ $\Box$

From \eqref{e:PBRD}, $P_\sD^\sBR$ depends on the positions of the entries of $\hbx$,
thus we also write $P_\sD^\sBR$ as $P_\sD^\sBR(\hbx)$.

According to  \eqref{e:PBRD}, to compute $P_\sD^\sBR$, we need to know the positions of
$\hat{x}_i$ on $[\ell_i, u_i]$ for $i=1,\ldots, n$.
In practice this information is  unknown.
However,  it is easy to observe from \eqref{e:PBRD} that $P_\sD^\sBR$ has a lower bound which does not rely on
the position of $\hbx$ in the box.
\begin{corollary} \label{c:boundPBR}
Let $\hbx$ in \eqref{e:model} be a deterministic vector, then
\begin{align*}
\ P_\sD^\sBR\geq& \frac{\det(\R)}{(2\pi \sigma^2)^{n/2}}\\
&\times \int_{-1/2}^{1/2}\cdots \int_{-1/2}^{1/2}
  \exp\left(-\frac{\|\R\bxi\|^2_2}{2\sigma^2}\right)d\xi_1\cdots d\xi_n,
\end{align*}
where the lower bound is reached if and only if $\ell_i < \hx_i <u_i$ for $i=1,\ldots,n$.
\end{corollary}

The lower bound is actually the success probability of the ordinary rounding detector, i.e., $P^\sOR$, see \cite[Th.\ 1]{WenTB16}.
It is easy to understand this.
In fact, the ordinary case can be regarded as a special situation of the box-constrained case:
 $\ell_i = -\infty$ and $u_i=\infty$,
thus, $\ell_i < \hx_i <u_i$  for $i=1,\ldots,n$.
Then,  the lower bound is reached and it is just $P^\sOR$.


The following theorem gives a formula for $P_\sR^\sBR$.
\begin{theorem}
 \label{t:PBRR}
Suppose that $\hbx$ in \eqref{e:model} is uniformly distributed over $\mathcal{B}$,
and $\hbx$ and $\v$ are independent, then
\begin{align}
\label{e:PBRR}
P_\sR^\sBR=\frac{1}{\prod_{i=1}^n(u_i-\ell_i+1)}\sum_{\forall \bar{\x}\in\mathcal{B}}
P_\sD^\sBR(\bar{\x}).
\end{align}
\end{theorem}
{\bf Proof.}  Notice that
\begin{align*}
\Pr(\x^\sBR=\hbx) &= \sum_{\forall \bar{\x}\in\mathcal{B}} \Pr(\x^\sBR = \hbx| \hbx=\bbx) \Pr(\hbx=\bbx)\\
&= \sum_{\forall \bar{\x}\in\mathcal{B}} \Pr(\x^\sBR =\bbx) \Pr(\hbx=\bbx).
\end{align*}

Since $\hbx$ is uniformly distributed over $\mathcal{B}$, for each $\bar{\x}\in \mathcal{B}$,
$$\Pr(\hbx=\bar{\x})=\frac{1}{\prod_{i=1}^n(u_i-\ell_i+1)}.$$
Therefore, \eqref{e:PBRR} holds.
\ \ $\Box$

Note that $P_\sR^\sBR$ can be computed, although the computational cost may be high
as the number of integer points in ${\cal B}$ can be large.

\subsection{Success probability of the box-constrained Babai detector}

In this subsection, we  give formulas for $P_\sD^\sBB$ and $P_\sR^\sBB$.
Since $P_\sD^\sBB$ depends on the position of $\hbx$ in the box $\mathcal{B}$,
we also give a lower bound on $P_\sD^\sBB$.

We first consider the deterministic situation.

\begin{theorem}
\label{t:PBBD}
Let $\hbx$ in \eqref{e:model} be a deterministic vector, then
\beq
P_\sD^\sBB=\prod_{i=1}^n\omega_i(r_{ii}),
\label{e:PBBD}
\eeq
where
\begin{equation}\label{e:omega}
\omega_i(r_{ii})=
\begin{cases}
\frac{1}{2}\big[1+\phi_\sigma(r_{ii})\big], &
       \mbox{ if }\  \hx_i=\ell_i  \mbox{ or }  \hx_i=u_i\\
\phi_\sigma(r_{ii}), &
 \mbox{ if }\  \ell_i<\hx_i< u_i
\end{cases}
\end{equation}
with
\begin{align}
\label{e:varphi}
\phi_\sigma(\zeta)
& :=\frac{2}{\sqrt{2\pi}}\int_{0}^{\frac{\zeta}{2\sigma}}\exp\big(-\frac{1}{2}t^2\big)dt    \nonumber\\
& = \frac{\zeta}{\sqrt{2\pi}\sigma} \int_{-\frac{1}{2}}^{\frac{1}{2}} \exp\big(\! -\frac{\zeta^2 t^2}{2\sigma^2}\big)dt.
\end{align}
\end{theorem}

{\bf Proof.}
From \eqref{e:modelqr},  for $i=n,  \ldots, 1$,
\[
\ty_i = r_{ii} \hx_i + \sum_{j=i+1}^n r_{ij}\hx_j + \tv_i.
\]
Then, using \eqref{e:Babai}, we obtain
\beq
c_i = \hx_i + \sum_{j=i+1}^n \frac{r_{ij}}{r_{ii}} (\hx_j-x_j^\sBB) + \frac{\tv_i}{r_{ii}}.
\label{e:ci}
\eeq
Therefore, if $x_{j}^\sBB=\hx_{j}$ for $j=i+1, \ldots, n$, then
\[
c_i   \sim\mathcal{N}(\hx_{i},  \sigma^2/r_{ii}^2).
\]

To simplify notation, denote events
$$
E_i = (x_{i}^\sBB=\hx_{i},  \ldots, x_n^\sBB = \hx_n), \quad i=1,\ldots, n.
$$
Then by the chain rule of conditional probabilities,
\beq\label{e:chain}
P_\sD^\sBB
 =\Pr(E_1) =\prod_{i=1}^{n}\Pr(x_i^\sBB=\hx_i|E_{i+1}),
\eeq
where $E_{n+1}$ is the sample space $\Omega$.

Now we consider $\Pr(x_i^\sBB=\hx_i|E_{i+1})$ for three different cases.

Case 1: $\hx_i=\ell_i$. In this case, by \eqref{e:Babai},
\begin{align*}
 & \Pr(x_i^\sBB=\hx_i \,|\, E_{i+1}) \\
=&  \Pr(c_i \leq \ell_i+1/2  \,|\, E_{i+1}) \\
=& \frac{1}{\sqrt{2\pi(\frac{\sigma}{r_{ii}})^2}}
   \int_{-\infty}^{\ell_i+\frac{1}{2}}\exp\Big(-\frac{(t-\ell_i)^2}{2(\frac{\sigma}{r_{ii}})^2}\Big)dt   \\
 =& \frac{1}{\sqrt{2\pi}}\int_{-\infty}^{\frac{r_{ii}}{2\sigma}}\exp\big(-\frac{t^2}{2}\big)dt
=   \frac{1}{2}[1+\phi_\sigma(r_{ii})].
\end{align*}

Case 2: $\ell_i<\hx_i<u_i$. In this case, by \eqref{e:Babai},
\begin{align*}
 & \Pr(x_i^\sBB=\hx_i  \,|\, E_{i+1}) \\
 =&  \Pr(  \hx_i -1/2\leq  c_i \leq  \hx_i +1/2  \,|\, E_{i+1}) \\
=& \frac{1}{\sqrt{2\pi(\frac{\sigma}{r_{ii}})^2}}
\int^{\hx_i+\frac{1}{2}}_{\hx_i-\frac{1}{2}}\exp\Big(-\frac{(t-\hx_i)^2}{2(\frac{\sigma}{r_{ii}})^2}\Big)dt
=\phi_\sigma(r_{ii}).
\end{align*}

Case 3: $\hx_i=u_i$. In this case, by \eqref{e:Babai},
\begin{align*}
 & \Pr(x_i^\sBB=\hx_i \,|\, E_{i+1}) \\
=&  \Pr(u_i-1/2 \leq c_i  \,|\, E_{i+1}) \\
=& \frac{1}{\sqrt{2\pi(\frac{\sigma}{r_{ii}})^2}}
   \int_{u_i-\frac{1}{2}}^\infty \exp\Big(-\frac{(t-u_i)^2}{2(\frac{\sigma}{r_{ii}})^2}\Big)dt \\
=& \frac{1}{2}[1+\phi_\sigma(r_{ii})].
\end{align*}
Therefore, from \eqref{e:chain}, this theorem holds. \ \ $\Box$
\medskip

The formula \eqref{e:PBBD} was originally given in the MSc thesis \cite{Han12},
supervised by the second author of this paper.
The proof given here is easier to follow than that given in \cite{Han12}.
Note that the main idea of its proof is similar to that of \cite[Th. 1]{WenC17}.


From Theorem \ref{t:PBBD}, similarly to $P_\sD^\sBR$,
to compute $P_\sD^\sBB$, we need to know the locations of $\hat{x}_i$ in the box $\mathcal{B}$.
But, these information is usually unknown in practice.
However, by \eqref{e:PBBD} and \eqref{e:omega}, the following corollary which gives
a lower bound and an upper bound on $P_\sD^\sBB$, that do not need priori information on  $\hbx$,
clearly holds.

\begin{corollary} \label{c:bound}
Let $\hbx$ in \eqref{e:model} be a deterministic vector, then
\begin{equation} \label{e:BBLUBD}
\prod_{i=1}^n \phi_\sigma(r_{ii}) \leq P_\sD^\sBB \leq \frac{1}{2^n} \prod_{i=1}^n (1+\phi_\sigma(r_{ii})),
\end{equation}
where the lower bound is reached if and only if $\ell_i < \hx_i <u_i$ for $i=1,\ldots,n$,
and the upper  bound is reached if and only if $\hx_i =\ell_i$ or $\hx_i=u_i$ for $i=1,\ldots,n$.
\end{corollary}


%
%

The lower bound given in the corollary  is actually the success probability of the ordinary Babai detector,
see \cite[eq. (11)]{ChaWX13}.


%
%

For the random situation, we have the following theorem for computing $P_\sR^\sBB$, see  \cite[Th. 1]{WenC17}.
\begin{theorem}
\label{t:PBBR}
Suppose that  $\hbx$ in \eqref{e:model} is uniformly distributed over $\mathcal{B}$,
and $\hbx$ and $\tilde{\v}$ are independent, then
\beq
P_\sR^\sBB=\prod_{i=1}^n \Big[\frac{1}{u_i-\ell_i+1}+\frac{u_i-\ell_i}{u_i-\ell_i+1}\phi_\sigma(r_{ii})\Big],
\label{e:PBBR}
\eeq
where  $\phi_\sigma(\zeta)$ is defined in \eqref{e:varphi}.
\end{theorem}

\section{Relationship between $P^\sBR$ and $P^\sBB$}
\label{s:PRPBrelation}

It has been showed in \cite[eq. (20)]{Teu98a} that the success probability of the ordinary
rounding detector cannot be larger than that of the ordinary Babai detector.
For the box-constrained case,  in this section, we will show that
the conclusion does not hold any more when the parameter vector is deterministic
while it still holds when the parameter vector is uniformly distributed.

Simulations show that, in general, $P_\sD^\sBR\leq P_\sD^\sBB$.
However, the following example shows that for the deterministic case
it is possible that $P_\sD^\sBR>P_\sD^\sBB$.

\begin{example}\label{ex:rounding}
Let $\sigma=1$, $\R=\bmx 2& -1\\ 0 & 1\emx$,  $\hx_{1}=\ell_{1}$ and $\hx_{2}=\ell_{2}$.
Then, by Theorems and \ref{t:PBRD} and \ref{t:PBBD}, we have
\[
P_\sD^\sBB=\frac{1}{4}(1+\phi_1(1))(1+\phi_1(2))=0.5818
\]
and
\[
P_\sD^\sBR=\frac{2}{2\pi}\int_{-\infty}^{1/2}\int_{-\infty}^{1/2}\exp(-\frac{1}{2}\|\R\bxi\|^2_2)d\xi_2d\xi_1=0.6192.
\]
Thus, $P_\sD^\sBR>P_\sD^\sBB$.
\end{example}

However, if $\hbx$  is uniformly distributed over $\mathcal{B}$, then $P_\sD^\sBR<P_\sD^\sBB$.
To prove this, we introduce  a lemma.



\begin{lemma}
 \label{l:integralineq2}
Suppose that $a>0$ and $s_2,\ldots, s_{n}$ are intervals, then for any $\sigma>0$,
 \begin{align}
 \label{e:integralineq2}
&\int_{s_{n}}\cdots \int_{s_{2}}\int_{-a}^a
\exp\left(-\frac{\|\R\bxi\|^2}{2\sigma^2}\right)d\xi_{1}\cdots d\xi_n\nonumber\\
\leq&\int_{-a}^a\exp\left(-\frac{r_{11}^2t^2}{2\sigma^2}\right)dt\nonumber\\
& \times \int_{s_{n}}\cdots \int_{s_{2}}
\exp\left(-\frac{\|\R_{2:n,2:n}\bxi_{2:n}\|^2}{2\sigma^2}\right)d\xi_{2}\cdots d\xi_{n}.
\end{align}
\end{lemma}

{\bf Proof.} We prove \eqref{e:integralineq2} by changing variables in the integral.

Let
\[
\T =\bmx
1& - \frac{1}{r_{11}} \R_{1,2:n} \\
\0& \I_{n-1}
\emx.
\]
Then
$$
\R\T = \bmx r_{11} & 0 \\ \0 & \R_{2:n,2:n} \emx.
$$
Define $\bxi= \T \boldsymbol{\eta}$, then  with
\[
s_1:=\big\{\eta_1|-a\leq \eta_1 - \sum_{j=2}^n\frac{r_{1j}}{r_{11}}\eta_j\leq a, \eta_2\in s_2,\ldots, \eta_n\in s_n \big\}.
\]
we have
\begin{align*}
&\int_{s_{n}}\cdots \int_{s_{2}}\int_{-a}^a
\exp\left(-\frac{\|\R\bxi\|^2}{2\sigma^2}\right)d\xi_{1}\cdots d\xi_n\nonumber\\
=&\int_{s_{n}}\cdots \int_{s_1}
\exp\left(-\frac{r_{11}^2\eta^2_1+\|\R_{2:n,2:n}\boldsymbol{\eta}_{2:n}\|^2}{2\sigma^2}\right)
d\eta_{1}\cdots d\eta_n\nonumber\\
=&\int_{s_{n}}\cdots \int_{s_{2}} \left(\int_{s_1}\exp\left(-\frac{r_{11}^2\eta^2_1}{2\sigma^2}\right)d\eta_1\right) \nonumber\\
&\times
\exp\left(-\frac{\|\R_{2:n,2:n}\boldsymbol{\eta}_{2:n}\|^2}{2\sigma^2}\right) d\eta_{2}\cdots d\eta_n.
\end{align*}

According to \cite[eq.\ (68)]{WenC17}, we have
\beq
 \label{e:Gaussianine}
\int_{s_1}\exp\left(-\frac{r_{11}^2\eta^2_1}{2\sigma^2}\right)d\eta_1
\leq \int_{-a}^{a}\exp\left(-\frac{r_{11}^2t^2}{2\sigma^2}\right)dt.
\eeq
Thus, \eqref{e:integralineq2} holds.
Note that \eqref{e:Gaussianine} can be easily  observed from the graph of the density function of the normally distributed random variable with 0 mean.
$\Box$

Here we make a remark.
It is easy to see from the lemma that if $a_i >0$ for $i=1,\ldots,n$, then
\begin{align*}
&\int_{-a_n}^{a_n}\cdots \int_{-a_1}^{a_1}
\exp\left(-\frac{\|\R\bxi\|^2_2}{2\sigma^2}\right)d\xi_1\cdots d\xi_n\\
\leq&\prod_{i=1}^n \int_{-a_i}^{a_i}\exp\left(-\frac{r_{ii}^2t^2}{2\sigma^2}\right)dt.
\end{align*}
If $a_i=1/2$ for $i=1,\ldots,n$, the above inequality leads to \cite[eq. (20)]{Teu98a}, which shows
that  the success probability of ordinary rounding detectors cannot be larger than
the success probability of ordinary Babai detectors,
but our proof is much simpler.

The following theorem characterizes the relationship between $P_\sR^\sBR$ and $P_\sR^\sBB$.
\begin{theorem}
 \label{t:PBRPBB}
Suppose that $\hbx$ is uniformly distributed over $\mathcal{B}$ and $\hbx$ and $\v$ are independent, then
\beq
\label{e:PBRPBB}
P_\sR^\sBR \leq P_\sR^\sBB.
\eeq
\end{theorem}

{\bf Proof.}
We prove \eqref{e:PBRPBB} by induction. Clearly, \eqref{e:PBRPBB} holds if $n=1$
since $\x^\sBR=\x^\sBB$ in this case.

In the following, we assume that \eqref{e:PBRPBB} holds for $n=k$ for any positive integer $k$,
then by induction, we show that  it also holds for $n=k+1$.

Denote
$
\mathcal{\tilde{B}}=\{\tilde{\x} \in \mathbb{Z}^{k}:\l_{2:k+1} \leq \tilde{\x} \leq \u_{2:k+1}\},
$
where $\l$ and $\u$ are defined in  \eqref{e:box}.
Write $ \bbx = \bmx \bx_1 \\ \tilde{\x} \emx$.
Then,
\begin{align*}
\sum_{\forall \bar{\x}\in\mathcal{B}}P_\sD^\sBR(\bar{\x})
=&\sum_{\forall \tilde{\x}\in\mathcal{\tilde{B}}}P_\sD^\sBR\left(\bmx \ell_1 \\ \tilde{\x}\emx\right)
+\sum_{\forall \tilde{\x}\in\mathcal{\tilde{B}}}P_\sD^\sBR\left(\bmx u_1 \\ \tilde{\x}\emx\right) \\
&+\sum_{i=1}^{u_1-\ell_1-1}\sum_{\forall \tilde{\x}\in\mathcal{\tilde{B}}}
P_\sD^\sBR \left(\bmx \ell_1+i \\ \tilde{\x}\emx\right).
\end{align*}

By Theorem \ref{t:PBRD}, Lemma \ref{l:integralineq2} and \eqref{e:varphi}, we have
\begin{align*}
&\sum_{\forall \tilde{\x}\in\mathcal{\tilde{B}}}P_\sD^\sBR\left(\bmx \ell_1 \\ \tilde{\x}\emx\right)
+\sum_{\forall \tilde{\x}\in\mathcal{\tilde{B}}}P_\sD^\sBR\left(\bmx u_1 \\ \tilde{\x}\emx\right)  \\
\leq& \left[\frac{r_{11}}{\sqrt{2\pi\sigma^2}}\int_{-\infty}^{\infty}\exp\left(-\frac{r_{11}^2t^2}{2\sigma^2}\right)dt
\right.\\
&\left.+\frac{r_{11}}{\sqrt{2\pi\sigma^2}}\int_{-\frac{1}{2}}^{\frac{1}{2}}\exp\left(-\frac{r_{11}^2t^2}{2\sigma^2}\right)dt
\right]
\sum_{\forall \tilde{\x}\in\mathcal{\tilde{B}}}P_\sD^\sBR(\tilde{\x})\\
=&(1+\phi_\sigma(r_{11}))\sum_{\forall \tilde{\x}\in\mathcal{\tilde{B}}}P_\sD^\sBR(\tilde{\x}).
\end{align*}

Similarly, we obtain
\begin{align*}
&\sum_{i=1}^{u_1-\ell_1-1}\sum_{\forall \tilde{\x}\in\mathcal{\tilde{B}}}
P_\sD^\sBR \left(\bmx \ell_1+i \\ \tilde{\x}\emx\right) \\
\leq & \ (u_1-\ell_1-1)\frac{r_{11}}{\sqrt{2\pi\sigma^2}}\int_{-\frac{1}{2}}^{\frac{1}{2}}\exp\left(-\frac{r_{11}^2t^2}{2\sigma^2}\right)dt \\
& \times \sum_{\forall \tilde{\x}\in\mathcal{\tilde{B}}}P_\sD^\sBR(\tilde{\x})\\
=& \ (u_1-\ell_1-1)\phi_\sigma(r_{11})\sum_{\forall \tilde{\x}\in\mathcal{\tilde{B}}}P_\sD^\sBR(\tilde{\x}).
\end{align*}
Therefore, by the above inequalities, we obtain
\begin{align*}
\sum_{\forall \bar{\x}\in\mathcal{B}}P_\sD^\sBR(\bar{\x})
\leq [1+(u_1-\ell_1)\phi_\sigma(r_{11})]
\sum_{\forall \tilde{\x}\in\mathcal{\tilde{B}}}P_\sD^\sBR(\tilde{\x}).
\end{align*}
Then, by Theorem \ref{t:PBRR} and Theorem \ref{t:PBBR}, we have
\begin{align*}
P_\sR^\sBR
\leq&\ \Big[\frac{1}{u_1-\ell_1+1}+\frac{(u_1-\ell_1)}{u_1-\ell_1+1}\phi_\sigma(r_{11})\Big]\\
&\times\frac{1}{\prod_{i=2}^n(u_i-\ell_i+1)}
\sum_{\forall \tilde{\x}\in\mathcal{\tilde{B}}}P_\sD^\sBR(\tilde{\x})\\
\leq &\ \Big[\frac{1}{u_1-\ell_1+1}+\frac{(u_1-\ell_1)}{u_1-\ell_1+1}\phi_\sigma(r_{11})\Big]\\
&\times\prod_{i=2}^n \Big[\frac{1}{u_i-\ell_i+1}+\frac{u_i-\ell_i}{u_i-\ell_i+1}\phi_\sigma(r_{ii})\Big]  \\
= &\ P_\sR^\sBB
\end{align*}
where the second inequality follows from the induction hypothesis.
$\Box$

\section{Numerical Simulations}
\label{s:sim}

In this section, we do numerical tests to illustrate Theorems \ref{t:PBRR}, \ref{t:PBBR} and
\ref{t:PBRPBB}. We let $n=8$ and generated 100 different $\A$'s by letting $\A=\text{randn}(n)$.
For each generated $\A$, we generated 1000 $\hbx$'s with each of them being uniformly
distributed over $\mathcal{B}=[0,3]^n$ and 1000 $\v$'s with $\v=\sigma\,\text{randn}(n,1)$,
 where $\sigma=0.05:0.05:0.4$.
For each generated $\A$, we use Theorems \ref{t:PBRR} and \ref{t:PBBR} to compute
$P_\sR^\sBR$ and $P_\sR^\sBB$, take their average values and denote them as
``Theo. $P_\sR^\sBR$" and ``Theo. $P_\sR^\sBB$", respectively.
We compute the experimental $P_\sR^\sBR$ and $P_\sR^\sBB$
which are the number of events $\x^\sBR=\hbx$ and $\x^\sBB=\hbx$
divided by $10^5$, and respectively denote them as
``Exp. $P_\sR^\sBR$" and ``Exp. $P_\sR^\sBB$".

\begin{figure}[!htbp]
\centering
\includegraphics[width=3.2in]{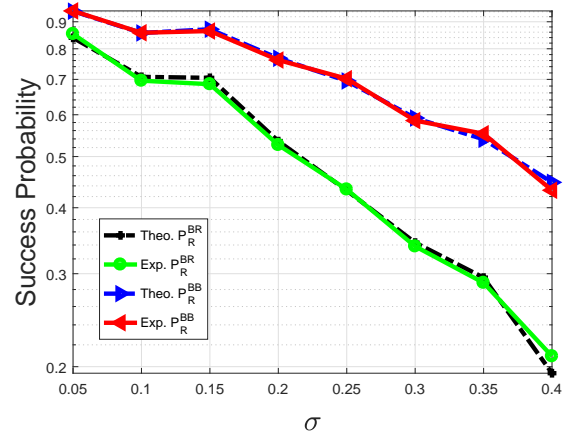}
\caption{Average success probabilities versus $\sigma$ for $n=8$ and $\mathcal{B}=[0,3]^n$}
\label{fig:sp}
\end{figure}

Figure \ref{fig:sp} shows the average success probabilities of the box-constrained rounding
and Babai detects versus $\sigma$ for $n=8$ and $\mathcal{B}=[0,3]^n$.
From Figure \ref{fig:sp}, one can see that the experimental success probabilities
of these two detectors are closely consistent with the success probabilities computed via
Theorems \ref{t:PBRR} and \ref{t:PBBR}, and Theorem \ref{t:PBRPBB} holds.

\section{Summary} \label{s:sum}

In this paper, we investigated the success probability of box-constrained rounding detectors
$\x^\sBR$ and box-constrained Babai detectors $\x^\sBB$, and studied their relationship.
We first proposed formulas for the success probability  $P_\sD^\sBR$ and $P_\sR^\sBR$
for $\x^\sBR$, and the success probability $P_\sD^\sBB$ for $\x^\sBB$.
Then, we gave an example which shows that $P_\sD^\sBR$ may strictly larger than $P_\sD^\sBB$.
Finally, we rigorously showed that $P_\sR^\sBR\leq P_\sR^\sBB$ always holds
if $\hbx$ is uniformly distributed over the constraint box $\mathcal{B}$.

\bibliographystyle{IEEEtran}
\bibliography{ref}
\end{document}